\documentclass[draftclsnofoot, onecolumn]{IEEEtran}
\usepackage{amsmath,amsfonts}
\usepackage{algorithmic}
\usepackage{algorithm}
\usepackage{array}
\usepackage[caption=false,font=normalsize,labelfont=sf,textfont=sf]{subfig}
\usepackage{textcomp}
\usepackage{stfloats}
\usepackage{url}
\usepackage{verbatim}
\usepackage{graphicx}
\usepackage{cite}

\begin{document}
\title{Atomic superheterodyne receiver Sensitivity estimation based on homodyne readout}

\author{Shanchi Wu, Chen Gong, Rui Ni
	\thanks{Shanchi Wu, Chen Gong are with University of Science and Technology of China, Email address: wsc0807@mail.ustc.edu.cn, cgong821@ustc.edu.cn.}
	\thanks{Rui Ni is with Huawei Technology, Email address: raney.nirui@huawei.com.}
	}

\maketitle

\begin{abstract} 
The electric field measurement sensitivity based on the Rydberg atomic vapor cell has great theoretical advantages over traditional dipole antennas. We combine the Rydberg atomic heterodyne receiver and the Mach-Zehnder interferometer (MZI) with high phase detection sensitivity to evaluate the system reception sensitivity based on the transmitted laser phase shift. We conduct a theoretical investigation into the impacts of local microwave electric field frequency detuning, and laser frequency detuning on enhancing the sensitivity of heterodyne Rydberg atomic receiver based on MZI. To optimize the output signal’s amplitude given the input microwave signal, we derive the steady-state solutions of the atomic density matrix. Numerical results show that laser frequency detuning and local microwave electric field frequency detuning can improve the system detection sensitivity, which can help the system achieve extra sensitivity gain. It also shows that the phase-based readout scheme of heterodyne Rydberg atomic receiver based on MZI can achieve better sensitivity than the intensity-based readout scheme of heterodyne Rydberg atomic receiver.
\end{abstract}

\begin{IEEEkeywords}
	Rydberg atom, phase, homodyne readout, frequency detuning, sensitivity optimization.
\end{IEEEkeywords}

\section{Introduction}
\IEEEPARstart{R}{ydberg}  atoms show extremely strong microwave transition electric dipole moments, which are sensitive to external electromagnetic fields. At room temperature, electromagnetic fields can be measured with high sensitivity and precision using atomic quantum coherence effects. Due to their small size, high sensitivity, and wide operating frequency band, Rydberg atomic electric field sensors offer great potential for the detection of radio frequency signals.

Utilizing EIT (electromagnetically induced transparency) spectroscopy, a Rydberg atomic sensor with sensitivity 30 uV$\cdot$cm$^{-1}$Hz$^{-1/2}$ and minimum detectable electric field intensity of $8$ uV/cm has been demonstrated \cite{sedlacek2012microwave}. Rydberg atomic sensors' sensitivity can be raised to $2$ uV$\cdot$cm$^{-1}$Hz$^{-1/2}$ using balanced homodyne detection and frequency modulation methods based on optical interferometers \cite{kumar2017atom, kumar2017rydberg}. Combining the Rydberg atomic sensor with the traditional superheterodyne approach results in a significant increase in sensitivity. Such method introduces a local microwave signal and the phase and frequency measurement of microwaves can be realized by the Rydberg atomic sensor, improving the sensitivity of microwave electric field detection to $55$ nV$\cdot$cm$^{-1}$Hz$^{-1/2}$ \cite{jing2020atomic}. The setting of laser parameters also affects the electric field detection sensitivity of the Rydberg atomic sensor. Numerical simulations and experimental results show that the sensitivity of the Rydberg atomic sensor is related to the amplitude intensity of the two lasers and microwave electric. The microwave electric detection sensitivity can be improved to $12.5$ nV$\cdot$cm$^{-1}$Hz$^{-1/2}$ by detuning the coupling laser frequency, which is the highest sensitivity achieved in experiments so far \cite{cai2022sensitivity}. For applications in communication systems, improving the detection sensitivity and predicting the Rydberg atomic sensor performance under different parameter settings through theoretical analysis and numerical simulation are significant.

Rydberg atomic sensors' detection sensitivity is fundamentally constrained by quantum noise, which is orders of magnitude less prevalent than thermal noise. The coherence time of the atoms can be decreased in real systems by a variety of variables, which lowers the system's sensitivity to detection. The main relaxation mechanisms that affect Rydberg atomic sensor's sensitivity include collision broadening, transition time broadening \cite{sagle1996measurement}, power broadening \cite{sautenkov2017power} and laser linewidth \cite{lu1997electromagnetically, tanasittikosol2011microwave}. Atomic collisions with cavity walls as well as atomic collisions with one another cause the collision broadening effect. The collision broadening effect can be efficiently decreased by low atomic density. The atoms' mobility causes the transition time broadening effect, which depends on the temperature, atomic mass, and laser beam size. The power broadening originates from the instability of laser power, which can be squeezed by the power stabilization module in experiment. The laser linewidth is an inherent property of lasers, and narrow linewidth laser can improve the performance of Rydberg atomic sensors. Finding technical methods to eliminate or reduce the influence of these factors is crucial for future applications in communication system.

Recent progress in experiments show that reducing the laser linewidth and power fluctuation are feasible improvement methods. Experimental techniques with ultrastable cavities can achieve extremely narrow linewidth laser beam and high power stabilization, which can increase the system sensitivity to the level higher than $100$ nV$\cdot$cm$^{-1}$Hz$^{-1/2}$. The optical readout noise is the dominant factor limiting the detection sensitivity \cite{jing2020atomic, cai2022sensitivity}. Except for apparatus performance improvement, new detection and readout techniques are the options left. In the perspective of readout techniques, the experiment based on optical interferometer provides ideas for reducing readout noise. Methods that combine circulating cavity with compressed state light have the potential to further reduce the readout noise limit of the system. Existing experimental schemes adopt different detection techniques depending on the atomic response regime, such as AT splitting effect of atomic resonance \cite{sedlacek2012microwave}, AC Stark effect under off-resonant \cite{anderson2021self}, and the heterodyne readout technique with tuning capability \cite{gordon2019weak}. A continuously tunable electric field measurement based on the far off-resonant AC stark effect in a Rydberg atomic vapor cell shows comparable detection sensitivity with a resonant microwave-dressed Rydberg heterodyne receiver using the same system \cite{simons2021continuous}. Additional research about electric field detuning can make smooth transitions among these operating regimes, especially for modulated signals. The frequency modulation spectroscopy \cite{kumar2017rydberg}, lower noise photodetector, and the setting of operating state in the experiment are all sensitivity optimization methods worth trying.

In previous work, the intensity-based readout scheme has been widely studied \cite{jing2020atomic, cai2022sensitivity}, and the most sensitive Rydberg atomic heterodyne receiver receives the intensity signal of the transmitted laser through a photodetector\cite{cai2022sensitivity}. Although phase-based homodyne readout scheme also has been demonstrated\cite{kumar2017atom, vsibalic2018rydberg, xiao1995measurement}, the relationship between the transmitted laser phase shift and electric field strength remains to be studied. Considering that atomic media changes the phase of transmitted laser, and optical interferometer is highly sensitive to the phase shift, we construct the phase-based Rydberg atomic heterodyne receiver embedded in Mach-Zehnder interferometer. This scheme is expected to further improve the performance of atomic-based electric field sensing sensitivity. In this work, we further characterize the effects of local microwave, probing laser and coupling laser frequency detuning on the sensitivity of Rydberg atomic superheterodyne receiver based on homodyne readout. 

The reminder of this paper is organized as follows. In Section II, we introduce the system structure of proposed scheme, and outline the basic detection principles based on homodyne readout. In Section III, we optimize the frequency parameters based on atomic four-level model. We theoretically investigate the detection response to the local microwave, coupling laser and probing laser frequency detuning. In Section IV, we numerically characterize the system sensitivity variation. In Section V, we evaluate the electric field detection sensitivity based on different noise baselines. Finally in Section VI, we conclude this work.

\section{System Model}

\subsection{System Structure}
\begin{figure}[htbp]
	\centering
	\includegraphics[width = 0.7\textwidth]{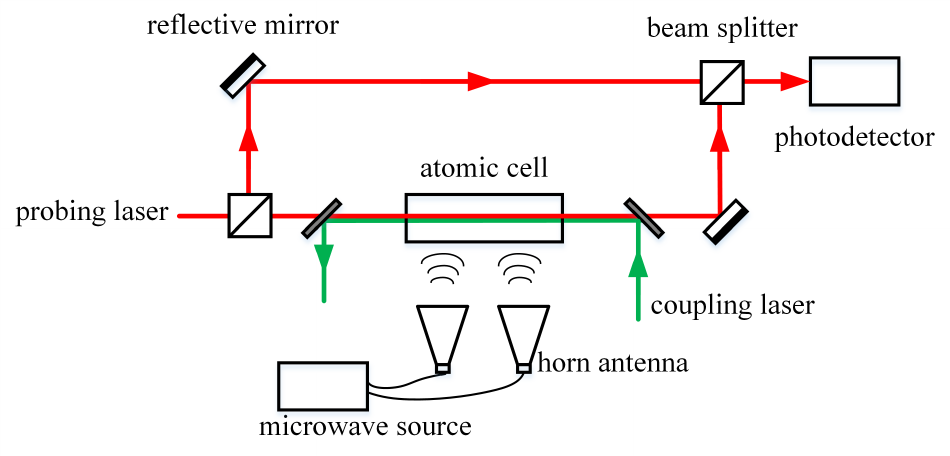}
	\caption{The diagram of the atomic heterodyne detection system based on homodyne readout.}
	\label{fig.system diagram}
\end{figure}

A schematic diagram of the Rydberg atomic superheterodyne receiver based on homodyne readout is shown in Fig. \ref{fig.system diagram}. It can be seen that compared with the Rydberg atomic heterodyne receiver based on intensity readout, the Rydberg atomic heterodyne receiver based on homodyne readout studied in this paper uses the MZI in the transmitted laser readout. The probing laser and coupling laser are locked on the desired wavelength and simultaneously excite the atoms in the atomic vapor cell. When the microwave signal radiates cross the atomic vapor cell, the atomic vapor cell changes the phase of transmitted probing laser, which is superimposed with the reference light through a beam splitter. Finally, the photodetector receives the mixed light for subsequent signal processing. The system response is sensitive to the direction of polarization. It has been demonstrated that the microwave electric field polarization can also be obtained by atomic system\cite{sedlacek2013atom}. In this work, we assume that the lasers and microwave signal have the same polarization.

\begin{figure}[htbp]
	\centering
	\includegraphics[width = 1\textwidth]{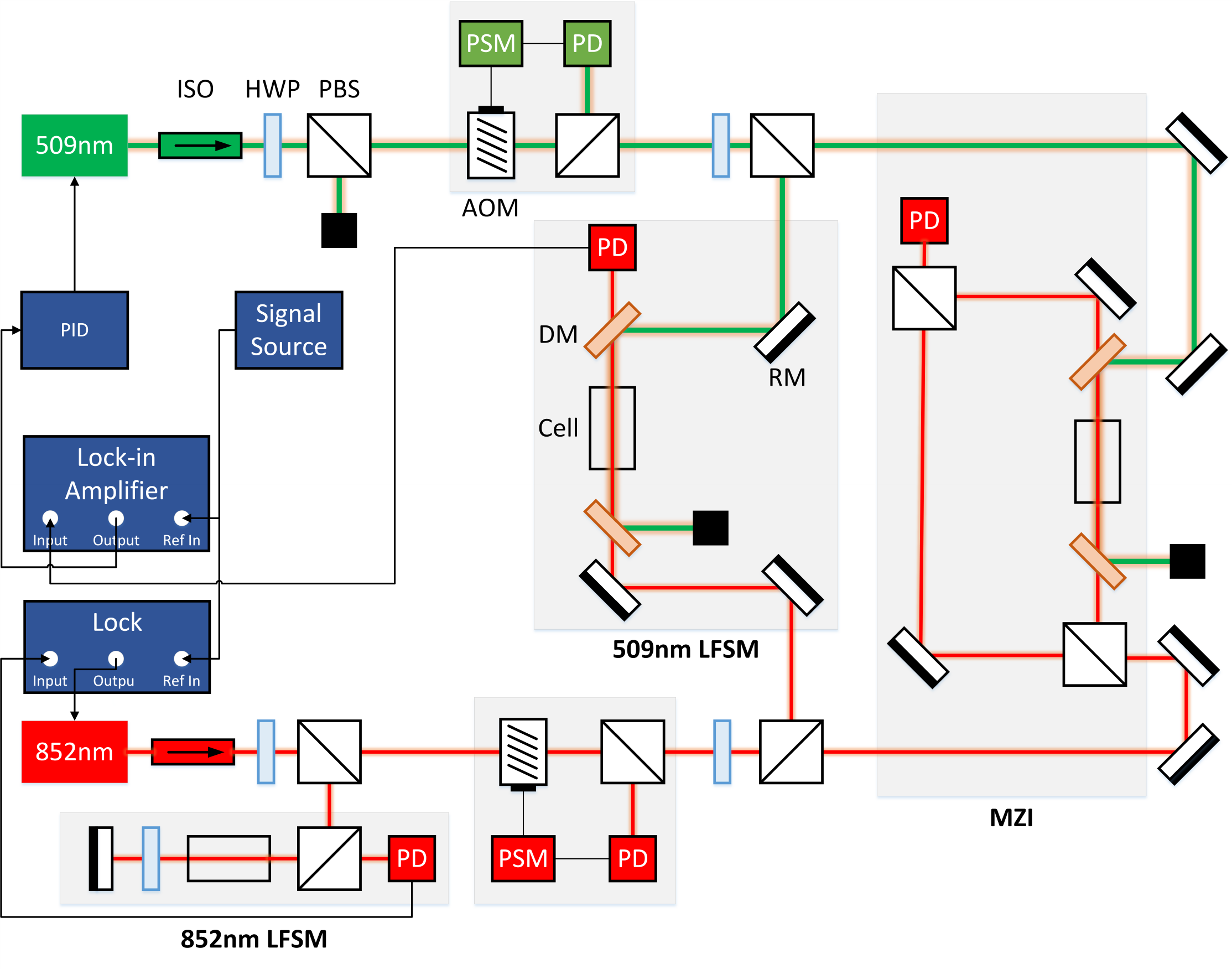}
	\caption{Overview of the system setup. We have used the following notations: (1) ISO: isolator, (2) HWP: half wave plate, (3) PBS: polarized beam splitter, (4) AOM: The acousto-optic modulators which shift the frequencies of the probing and coupling lights to atomic resonances, (5) PSM: power stabilization module, (6) PD: photodetector, (7) DM: dichroic mirror, (8) RM: reflective mirror, (9) LFSM: laser frequency stabilization module, (10) MZI: Mach-Zehnder interferometer.}
	\label{fig.system setup}
\end{figure}

In order to obtain a narrower laser linewidth and lower power fluctuations, the actual system is more complex. Fig. \ref{fig.system setup} shows the proposed system setup. The 852nm laser uses a frequency stabilization scheme based on atomic absorption spectroscopy, while the 509nm laser uses a frequency stabilization scheme based on EIT effect of atomic absorption spectra. The two lasers pass through the power stabilization module and then enter the detection atomic vapor cell to excite the atoms to the Rydberg state. Under the radiation of electric field, the atomic vapor cell changes the phase of the transmitted laser, which is received by the photodetector through a MZI. The frequency stabilization scheme can be replaced by other higher performance schemes, such as ultra-low expansion cavity scheme.

\subsection{Theoretical Model}
\begin{figure}[htbp]
	\centering
	\includegraphics[width = 2.5in]{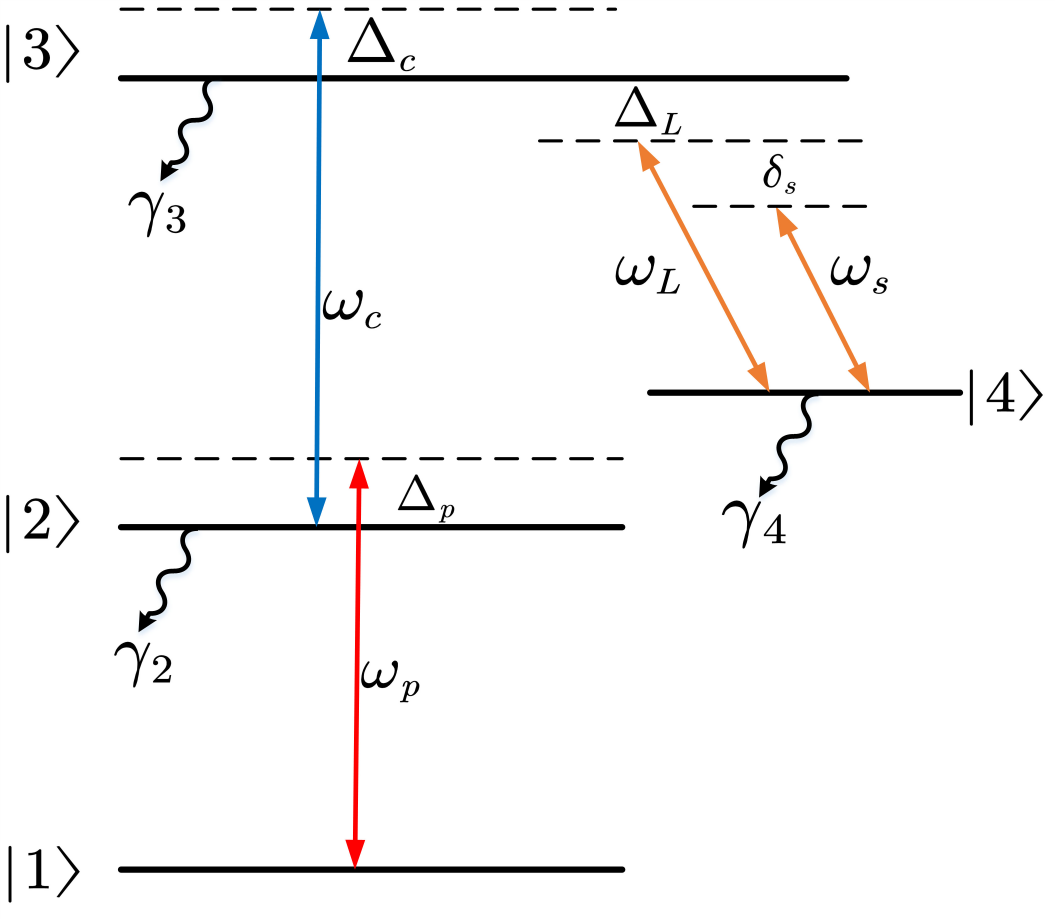}
	\caption{The diagram of four-level configuration.}
	\label{fig.level}
\end{figure}

The diagram for a typical atomic four-level structure is shown in Fig. \ref{fig.level}. The probing laser and coupling laser are tuned with detuning $\Delta _p$ and $\Delta _c$, respectively; and the local microwave electric field is tuned with detuning $\Delta _L$. Assume that the frequency and phase difference between local microwave electric field and signal microwave electric field is $\delta _s$ and $\phi_s$, respectively. The corresponding Hamiltonian in the rotating frame is given by
\begin{equation}
H=\frac{{\hbar }}{2}\left( \begin{matrix}
	0&		\Omega _p&		0&		0\\
	\Omega _p&		2\Delta _p&		\Omega _c&		0\\
	0&		\Omega _c&		2\left( \Delta _p+\Delta _c \right)&		\Omega _L+\Omega _se^{-iS\left( t \right)}\\
	0&		0&		\Omega _L+\Omega _se^{iS\left( t \right)}&		2\left( \Delta _p+\Delta _c+\Delta _L \right)\\
\end{matrix} \right) ,
\end{equation} 
where $\Omega _{p}$ and $\Omega _{c}$ are the corresponding Rabi frequencies for transitions $\left. |1 \right> \rightarrow \left. |2 \right>$ and $ \left. |2 \right> \rightarrow \left. |3 \right>$, respectively. The local microwave electric field and signal microwave electric field couple with Rydberg transition between state $ \left. |3 \right> $ and state $ \left. |4 \right>$, with Rabi frequencies $\Omega_L$ and $\Omega_s$, respectively. Let $\Omega =|\Omega _L+\Omega_se^{-iS\left( t \right)}|$, where $S\left( t \right) =2\pi \delta _st+\phi _s$ is the cumulative phase difference of the signal microwave relative to the local microwave.

Due to the relaxation effect caused by the spontaneous radiation, collision of atoms and transit relaxation effect, as well as the atoms repopulation, the complete Liouville equation for the rotating-frame density matrix of the system can be written as \cite{auzinsh2010optically}
\begin{equation}
	i{\hbar }\frac{d}{dt}\rho =\left[ H,\ \rho \right] -i{\hbar }\frac{1}{2}\left( \Gamma \rho +\rho \Gamma \right) +i{\hbar }\Lambda,
\end{equation}
where $\rho$ is the density matrix. Relaxation matrix $\Gamma$, which represents the relaxation rate of each state, is given by
\begin{equation}
	\Gamma =\left( \begin{matrix}
		\gamma&		0&		0&		0\\
		0&		\gamma +\gamma _2&		0&		0\\
		0&		0&		\gamma +\gamma _3+\gamma _c&		0\\
		0&		0&		0&		\gamma +\gamma _4\\
	\end{matrix} \right),
\end{equation}
where $\gamma_{2}$, $\gamma_{3}$ and $\gamma_{4}$ represent the spontaneous decay rates of atoms on the three high levels; $\gamma_c$ and $\gamma$ are the relaxation rates of the atoms collision and transit effect, respectively. We assume that each level undergoes the same relaxation rate $\gamma$ due to the exit of atoms from the laser beam.

In addition, the atoms that spontaneously decay from the upper states also repopulate the lower sates. In this work, we only consider the decay paths shown in Fig. \ref{fig.level}. Repopulation matrix $\Lambda$ is given by
\begin{equation}
	\Lambda =\left( \begin{matrix}
		\gamma +\gamma _2\rho _{22}+\gamma _4\rho _{44}&		0&		0&		0\\
		0&		\gamma _3\rho _{33}&		0&		0\\
		0&		0&		0&		0\\
		0&		0&		0&		0\\
	\end{matrix} \right).
\end{equation}
The dynamics equation of the atomic system can be expressed as
\begin{equation}\label{eq.basic}
	\frac{d}{dt}\rho =-\frac{i}{{\hbar }}\left[ H,\rho \right] +L,
\end{equation}
where $L=-\frac{1}{2}\left( \Gamma \rho +\rho \Gamma \right) + \Lambda$ represents the Lindbladian operator that defines the relaxation progress in the system, which matrix form is given by
\begin{equation}
	L= \left( \begin{matrix}
		\gamma +\gamma _2\rho _{22}+\gamma _4\rho _{44}&		-\frac{\gamma _2+2\gamma}{2}\rho _{12}&		-\frac{\gamma _3+\gamma _c+2\gamma}{2}\rho _{13}&		-\frac{\gamma _4+2\gamma}{2}\rho _{14}\\
		-\frac{\gamma _2+2\gamma}{2}\rho _{21}&		\left( \gamma _3+\gamma _c+\gamma \right) \rho _{33}-\left( \gamma _2+\gamma \right) \rho _{22}&		-\left( \frac{\gamma _{23}+\gamma _c+2\gamma}{2} \right) \rho _{23}&		-\left( \frac{\gamma _{24}+2\gamma}{2} \right) \rho _{24}\\
		-\frac{\gamma _3+\gamma _c+2\gamma}{2}\rho _{31}&		-\left( \frac{\gamma _{23}+\gamma _c+2\gamma}{2} \right) \rho _{32}&		-\left( \gamma _3+\gamma _c+\gamma \right) \rho _{33}&		-\left( \frac{\gamma _{34}+\gamma _c+2\gamma}{2} \right) \rho _{34}\\
		-\frac{\gamma _4+2\gamma}{2}\rho _{41}&		-\left( \frac{\gamma _{24}+2\gamma}{2} \right) \rho _{42}&		-\left( \frac{\gamma _{34}+\gamma _c+2\gamma}{2} \right) \rho _{43}&		-\left( \gamma _4+\gamma \right) \rho _{44}\\
	\end{matrix} \right) .
\end{equation}

\subsection{Principle}
We begin with analytical derivations based on cold atoms. Within the adiabatic approximation, the probe laser transmission associated with the instantaneous steady state is written in terms of the imaginary component of the susceptibility as $ P\left( t \right) =P_ie^{-k_pL\Im \left( \chi \right)} $. Here, $P_i$ is the incident light power, $L$ is the length of atomic vapor cell and $ k_p=2\pi /\lambda _p $ is the probing laser wavevector \cite{jing2020atomic}. The linear susceptibility can be given by
\begin{equation}
	\chi =-\frac{2N_0\mu _{21}^{2}}{{\hbar }\varepsilon _0\Omega _p}\rho _{21},
\end{equation}
where $N_0$ is the total density of atoms, $\mu_{21}$ is the dipole moment of transition $|1\rangle \rightarrow |2\rangle$ and $\epsilon_0$ is the permittivity in vacuum, and $\rho_{21}$ is the density matrix element that describes the atomic coherence between the ground and first excited state and is found by solving Eq. \ref{eq.basic} in steady state.

The presence of the atomic ensemble causes absorption and phase shifts on the probe light that depend on the density matrix via the differential equations \cite{meyer2021optimal}
\begin{equation}
	\frac{d\Omega _p}{dz}=\frac{k_pN_0\mu _{21}^{2}}{2{\hbar }\varepsilon _0}\Im \left( \rho _{21} \right) ,
\end{equation}
and
\begin{equation}
	\frac{d\phi}{dz}=\frac{k_pN_0\mu _{21}^{2}}{2{\hbar }\varepsilon _0\Omega _p\left( z \right)}\Re \left( \rho _{21} \right).	
\end{equation}

We analyze the thin optical depth case and ignore the space dependency of $\Omega_p$. This yields the phase difference through the atomic vapor cell, 
\begin{equation}\label{eq.phase}
	\Delta \phi =k_pL\Re \left( \chi \right) /4.
\end{equation}
Then, the output electric field through the atomic medium is given by
\begin{equation}
	E_{out}=E_i\cdot e^{-k_pL\Im \left( \chi \right) /2}\cdot e^{i\Delta \phi}.
\end{equation}
In the framework of MZI, the electric field of signal arm (through the atomic vapor cell) and reference arm received by the photodetector are given by
\begin{equation}
	\begin{aligned}
		E_{\text{SIG}}&=E_1\cdot e^{ikn_0L_{\text{SIG}}}\cdot e^{-k_pL\Im \left( \chi \right) /2}\cdot e^{i\Delta \phi},\\
		E_{\text{REF}}&=E_2\cdot e^{ikn_0L_{\text{REF}}},\\
	\end{aligned}
\end{equation}
where $E_1$ and $L_{\text{SIG}}$ are the incident electric field and optical path length of signal arm, and $E_2$ and $L_{\text{REF}}$ are the incident electric field and optical path length of reference arm; $n_0$ is the refractive index. The differential signal of the balanced homodyne detectors (not shown in Fig. \ref{fig.system setup}) is given by \cite{xiao1995measurement}
\begin{equation}
	\Delta I_d\propto 2|E_{\text{REF}}||E_{\text{SIG}}|e^{-k_pL\Im \left( \chi \right) /2}\cos \left( \phi _{\text{REF}}+\Delta \phi \right).
\end{equation}

According to the signal of photodetector, the small phase shift $\Delta{\phi}$ can be inferred. The strength of signal electric field can be extracted based on Eq. (\ref{eq.phase}) and the relationship between $\Re({\chi})$ and $\Omega_s$, which is investigated in this work. Based on the steady-state solution of the system Liouville equation, we derive the expression of the linear susceptibility and phase shift, and then find the optimal values. 

\section{Parameters Optimization}\label{sec.frequency}
\subsection{Methodology}
In experimental systems, in order to reduce the interaction and collision rate between atoms, the atomic density and laser power are usually limited, although increasing the laser power can theoretically increase the signal-to-noise ratio of the output signal and improve the sensitivity. Therefore, in the current experimental setups, the detection laser power is usually at a weak level, and the coupled laser power is also limited. For Rydberg atom sensors using EIT readout, the optimal Rabi frequencies for transitions $|1\rangle \rightarrow |2\rangle$ and $|2\rangle \rightarrow |3\rangle$ depend on the decay rate of each state and relaxation rate caused by atom collision \cite{meyer2021optimal}. We analyze the influence of laser frequencies and local microwave electric field frequency detuning on the system sensitivity.

We set a detuning parameter $ \Delta$ (which can be $\Delta_p$, $\Delta _c$, or $\Delta _L$). Assuming $\gamma _3=\gamma _4=\gamma_c=\gamma=0$, the steady-state solution $\rho _{21}\left( \Delta \right)$ can be obtained based on Eq. (\ref{eq.basic}). Assuming $\Omega_s \ll \Omega_L$, the output optical power has the formalism as follows,
\begin{equation}\label{eq.output}
	P\left( t \right) \approx \bar{P}_0+\kappa \Omega _s\cos \left( 2\pi \delta _st+\phi _s \right),
\end{equation}
where $\kappa$ is a conversion coefficient \cite{jing2020atomic}. Similarly, the corresponding linear susceptibility real component of the atomic medium can be written as
\begin{equation}
\Re \left( \chi \left( t,\ \Delta \right) \right) \approx \chi _0\left( \Delta \right) +\chi _1\left( \Delta \right) \Omega _s\cos \left( 2\pi \delta _st+\phi _s \right),
\end{equation}
which has a time-invariant part $\chi _0\left( \Delta \right)$ and the coefficient of time variant part $\chi _1\left( \Delta \right)$. The first part contributes a constant phase shift, and the second part represents a phase variation that depends on the electric field strength. 

\begin{figure}[htbp]
	\centering
	\includegraphics[width = 3.5 in]{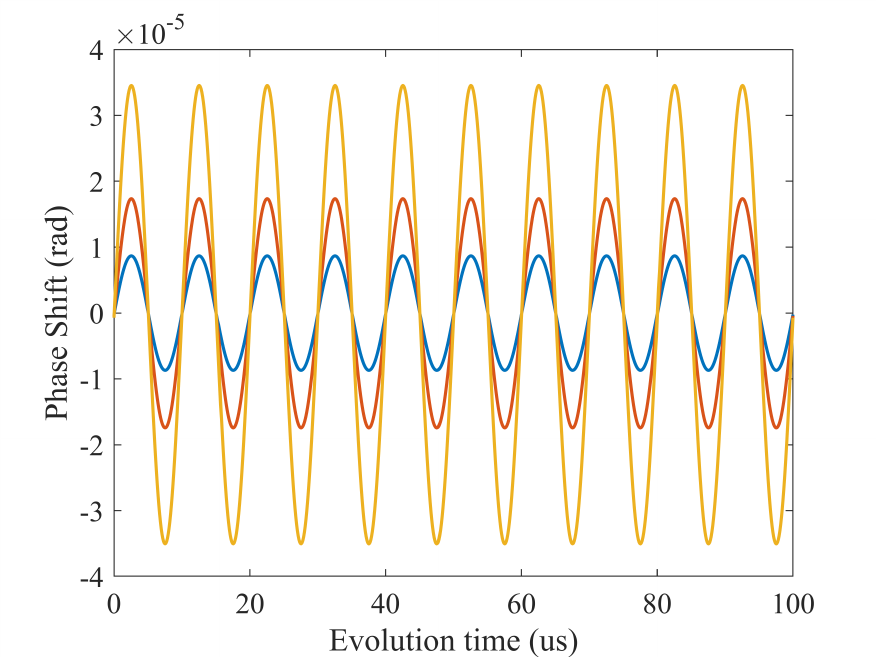}
	\caption{Transmitted laser phase shift $\Delta\phi$ evolution over time.}
	\label{fig.phase_shift}
\end{figure}

The probing laser transmission phase shift is associated with the real component of the susceptibility as
\begin{equation}
	\begin{aligned}\label{eq.delta_phi}
		\Delta \phi = k_p L \Re (\chi) / 4 = k_p L \chi_0(\Delta) + k_p L \chi_1 (\Delta) \Omega_s \cos(2\pi\delta_s + \phi_s).
	\end{aligned}
\end{equation}
Fig. \ref{fig.phase_shift} shows the transmitted laser phase shift evolution over time with different signal electric field strengths.

Since the amplitude of microwave signal is linearly dependent on its corresponding Rabi frequency $ \Omega_s$, we can optimize coefficient $ \chi _1\left( \Delta \right) $ in Eq. (\ref{eq.delta_phi}) to maximize the system sensitivity. It should be noted that the case of weak signal corresponds to small variation of probing laser transmission power. We analyze phase shift individually and do not consider laser power variation with time.

\subsection{Local Microwave Detuning}\label{sec.l_detuning}

Assuming that $\Delta _c=\Delta _p=\gamma _3=\gamma _4=0$, the steady-state solution $\rho_{21}(\Delta_{L})$ can be obtained according to Eq. (\ref{eq.basic}) as follows,
\begin{equation}\label{eq.rho21_L}
		\rho _{21}( \Delta _L )=-\frac{i\gamma _2\Omega _p\Omega ^4-2\Delta _L\Omega _p\Omega ^2\Omega _{c}^{2}}{\gamma _{2}^{2}\Omega ^4+4\Delta _{L}^{2}\left( \Omega _{c}^{2}+\Omega _{p}^{2} \right) ^2+2\Omega ^2\Omega _{p}^{2}\left( \Omega _{c}^{2}+\Omega _{p}^{2}+\Omega ^2 \right)}.\\
\end{equation}
The real component is given by
\begin{equation}
	\Re \left( \rho _{21}\left( \Delta _L \right) \right) =\frac{2\Delta _L\Omega _p\Omega ^2\Omega _{c}^{2}}{\gamma _{2}^{2}\Omega ^4+4\Delta _{L}^{2}\left( \Omega _{c}^{2}+\Omega _{p}^{2} \right) ^2+2\Omega ^2\Omega _{p}^{2}\left( \Omega _{c}^{2}+\Omega _{p}^{2}+\Omega ^2 \right)}.
\end{equation}

In the weak signal regime, $\Omega_s \ll \Omega_L$, the first order approximation of $\Omega_s / \Omega_L$ is given by,
\begin{equation}\label{eq.rho21_L_Im}
	\Re \left( \rho _{21}\left( \Delta _L \right) \right) \approx \frac{C_{L-DC}+C_{L-AC}\Omega _s\cos \left( S\left( t \right) \right)}{\left( \gamma _{2}^{2}\Omega _{L}^{4}+4\Delta _{L}^{2}\left( \Omega _{c}^{2}+\Omega _{p}^{2} \right) ^2+2\Omega _{L}^{2}\Omega _{p}^{2}\left( \Omega _{c}^{2}+\Omega _{p}^{2}+\Omega _{L}^{2} \right) \right) ^2},
\end{equation}
where $C_{L-DC}$ and $C_{L-AC}$ are as follows,
\begin{equation}
	C_{L-DC}=2\Delta _L\Omega _p\Omega _{c}^{2}\Omega _{L}^{2}\left( \gamma _{2}^{2}\Omega _{L}^{4}+4\Delta _{L}^{2}\left( \Omega _{p}^{2}+\Omega _{c}^{2} \right) ^2+2\Omega _{L}^{2}\Omega _{p}^{2}\left( \Omega _{p}^{2}+\Omega _{c}^{2}+\Omega _{L}^{2} \right) \right),
\end{equation}
and
\begin{equation}
	C_{L-AC}=4\Delta _L\Omega _{c}^{2}\Omega _p \Omega_L\left( 4\Delta _{L}^{2}\left( \Omega _{p}^{2}+\Omega _{c}^{2} \right) ^2-\Omega _{L}^{4}\left( \gamma _{2}^{2}+2\Omega _{p}^{2} \right) \right).
\end{equation}

The corresponding linear susceptibility of the atomic system can be written as
\begin{equation}
	\Re \left( \chi \left( t,\ \Delta _L \right) \right) =\chi _0\left( \Delta _L \right) +\chi _1\left( \Delta _L \right) \Omega _s\cos S\left( t \right),
\end{equation}
which has a time-invariant part,
\begin{equation}
	\chi _0\left( \Delta _L \right) =\frac{2N_0\mu _{12}^{2}}{{\hbar }\varepsilon _0}\frac{C_{L-DC}}{[\gamma _{2}^{2}\Omega _{L}^{4}+4\Delta _{L}^{2}\left( \Omega _{c}^{2}+\Omega _{p}^{2} \right) ^2+2\Omega _{L}^{2}\Omega _{p}^{2}\left( \Omega _{p}^{2}+\Omega _{c}^{2}+\Omega _{L}^{2} \right)]^2},
\end{equation}
and a time variant part coefficient,
\begin{equation}
	\chi _1\left( \Delta _L \right) =\frac{2N_0\mu _{12}^{2}}{{\hbar }\varepsilon _0}\frac{C_{L-AC}}{\left[ \gamma _{2}^{2}\Omega _{L}^{4}+4\Delta _{L}^{2}\left( \Omega _{c}^{2}+\Omega _{p}^{2} \right) ^2+2\Omega _{L}^{2}\Omega _{p}^{2}\left( \Omega _{c}^{2}+\Omega _{p}^{2}+\Omega _{L}^{2} \right) \right] ^2}.
\end{equation}

\textbf{(P1): Sensitivity Maximization in General Case via Local Microwave Detuning}
\begin{equation}\label{eq.opti1}
	\Delta _{L}^{*}=\underset{\Delta _L}{\text{argmax }}|\chi_{1}(\Delta_L)|.
\end{equation}

\subsection{Probing Laser Frequency Detuning}
Assuming that $\Delta _c=\Delta _L=\gamma _3=\gamma _4=\gamma_c=\gamma=0$, the steady-state solution $\rho_{21}(\Delta_{p})$ can be obtained according to Eq. (\ref{eq.basic}) as follows,
\begin{equation}\label{eq.rho21_p}
	\begin{aligned}
	\rho _{21}(\Delta_p) &=\frac{\Omega _p\left( 4\Delta _{p}^{2}-\Omega ^2 \right) \left( i\gamma _2\left( 4\Delta _{p}^{2}-\Omega ^2 \right) +2\Delta _p\left( 4\Delta _{p}^{2}-\Omega _{c}^{2}-\Omega ^2 \right) \right)}{C_P(\Omega)}, \\
	C_P(\Omega) &=64\Delta _{p}^{6}+\gamma _{2}^{2}\left( \Omega ^2-4\Delta _{p}^{2} \right) ^2 +4\Delta_{p}^{2}\left( \left( \Omega ^2+\Omega _{c}^{2} \right) ^2+2\Omega _{p}^{2}\left( \Omega _{p}^{2}+\Omega _{c}^{2}-2\Omega ^2 \right) \right) \\
	&-32\Delta _{p}^{4}\left( \Omega ^2+\Omega _{c}^{2}-\Omega _{p}^{2} \right) +2\Omega _{p}^{2}\Omega ^2\left( \Omega _{p}^{2}+\Omega _{c}^{2}+\Omega ^2 \right).
\end{aligned}
\end{equation}
The real component is given by
\begin{equation}
	\Re \left( \rho _{21}\left( \Delta _p \right) \right) =\frac{2\Delta _p\Omega _p\left( 4\Delta _{p}^{2}-\Omega ^2 \right) \left( 4\Delta _{p}^{2}-\Omega ^2-\Omega _{c}^{2} \right)}{C_P\left( \Omega \right)}.
\end{equation}

In weak signal regime, $\Omega_s \ll \Omega_L$, the first order approximation of $\Omega_s / \Omega_L$ is given by,
\begin{equation}\label{eq.rho21_p_Im}
	\Re \left( \rho _{21}\left( \Delta _p \right) \right) \approx -\frac{C_{P-DC}+C_{P-AC}\Omega _s\cos \left( S\left( t \right) \right)}{C_P\left( \Omega _L \right) ^2},
\end{equation}
where $C_{P-DC}$ and $C_{P-AC}$ are as follows,
\begin{equation}
	C_{P-DC}=2\Delta _p\Omega _p\left( 4\Delta _{p}^{2}-\Omega _{L}^{2} \right) \left( 4\Delta _{p}^{2}-\Omega _{c}^{2}-\Omega _{L}^{2} \right) C_P\left( \Omega _L \right),
\end{equation}
and
\begin{equation}
	\begin{aligned}
		C_{P-AC}&=4\Delta _p\Omega _p\Omega _L\left( 64\Delta _{p}^{6}\Omega _{c}^{2}-\gamma _{2}^{2}\left( 4\Delta _{p}^{2}-\Omega _{L}^{2} \right) ^2\Omega _{c}^{2}+2\Omega _{L}^{4}\Omega _{p}^{4}-32\Delta _{p}^{4}\left( \Omega _{c}^{4}+3\Omega _{p}^{4}+\Omega _{c}^{2}\left( \Omega _{L}^{2}+4\Omega _{p}^{2} \right) \right) \right) -\\
		&16\Delta _{p}^{3}\Omega _p\Omega _L\left( \Omega _{c}^{6}+4\Omega _{L}^{2}\Omega _{p}^{4}+2\Omega _{c}^{4}\left( \Omega _{L}^{2}+2\Omega _{p}^{2} \right) +\Omega _{c}^{2}\left( \Omega _{L}^{4}+8\Omega _{L}^{2}\Omega _{p}^{2}+4\Omega _{p}^{4} \right) \right).\\
	\end{aligned}
\end{equation}

The corresponding real component of linear susceptibility of the atomic system can be written as
\begin{equation}
	\Re \left( \chi \left( t,\ \Delta _p \right) \right) =\chi _0\left( \Delta _p \right) +\chi _1\left( \Delta _p \right) \Omega _s\cos S\left( t \right),
\end{equation}
which has a time-invariant part,
\begin{equation}\label{eq.chi0}
	\chi _0\left( \Delta _p \right) =\frac{2N_0\mu _{12}^{2}}{{\hbar }\varepsilon _0}\frac{C_{P-DC}}{C_P(\Omega_L)^2},
\end{equation}
and a time variant part coefficient,
\begin{equation}\label{eq.chi1}
	\chi _1\left( \Delta _p \right) =\frac{2N_0\mu _{12}^{2}}{{\hbar }\varepsilon _0}\frac{C_{P-AC}}{C_P(\Omega_{L}) ^2}.
\end{equation}

\textbf{(P2): Sensitivity Maximization in General Case via Probing Laser Frequency Detuning}
\begin{equation}\label{eq.opti2}
	\Delta _{p}^{*}=\underset{\Delta _p}{\text{argmax }}|\chi_{1}(\Delta_p)|.
\end{equation}

\subsection{Coupling Laser Frequency Detuning}

Assuming that $\Delta _p=\Delta _L=\gamma _3=\gamma _4=\gamma_c=\gamma=0$, the steady-state solution $\rho_{21}(\Delta_{c})$ can be obtained according to Eq. (\ref{eq.basic}) as follows,
	\begin{equation}\label{eq.rho21_c}
	\begin{aligned}
		\rho _{21}(\Delta_c) &=\frac{i\gamma _2\Omega _p\left( 4\Delta _{c}^{2}-\Omega ^2 \right) ^2-2\Omega _p\Omega _{c}^{2}\Delta _c\left( 4\Delta _{c}^{2}-\Omega ^2 \right)}{C_C(\Omega)}, \\
		C_C(\Omega) &=32\Delta _{c}^{4}\Omega _{p}^{2}+\gamma _{2}^{2}\left( \Omega ^2-4\Delta _{c}^{2} \right) ^2+2\Omega _{p}^{2}\Omega ^2\left( \Omega _{p}^{2}+\Omega _{c}^{2}+\Omega ^2 \right) \\
		& +4\Delta _{c}^{2}\left( \left( \Omega _{p}^{2}+\Omega _{c}^{2} \right) ^2+\Omega _{p}^{2}\left( \Omega _{p}^{2}-4\Omega ^2 \right) \right). 
	\end{aligned}
\end{equation}
The real component is given by
\begin{equation}
	\Re \left( \rho _{21}\left( \Delta _c \right) \right) =-\frac{2\Delta _c\Omega _p\Omega _{c}^{2}\left( 4\Delta _{c}^{2}-\Omega ^2 \right)}{C_C\left( \Omega \right)}.
\end{equation}

In weak signal regime, $\Omega_s \ll \Omega_L$, the first order approximation of $\Omega_s / \Omega_L$ is given by,
	\begin{equation}\label{eq.rho21_c_Im}
	\Re \left( \rho _{21}\left( \Delta _c \right) \right) =-\frac{C_{C-DC}+C_{C-AC}\Omega _s\cos \left( S\left( t \right) \right)}{C_C\left( \Omega _L \right) ^2},
\end{equation}
where $C_{C-DC}$ and $C_{C-AC}$ are as follows,
\begin{equation}
	C_{C-DC}=2\Delta _c\Omega _p\Omega _{c}^{2}\left( 4\Delta _{c}^{2}-\Omega _{L}^{2} \right) C_C\left( \Omega _L \right),
\end{equation}
and
\begin{equation}
	C_{C-AC}=4\Delta _c\Omega _p\Omega _{c}^{2}\Omega _L\left( \gamma _{2}^{2}\left( -4\Delta _{c}^{2}+\Omega _{L}^{2} \right) ^2+32\Delta _{c}^{4}\Omega _{p}^{2}+2\Omega _{L}^{4}\Omega _{p}^{2}-4\Delta _{c}^{2}\left( \Omega _{c}^{4}+4\Omega _{c}^{2}\Omega _{p}^{2}+4\Omega _{p}^{2}\left( \Omega _{L}^{2}+\Omega _{p}^{2} \right) \right) \right).
\end{equation}

The corresponding real component of linear susceptibility $\chi \left( t, \Delta_c \right)$ of the atomic system can be written as
\begin{equation}
	\Im\left( \chi \left( t, \Delta_c \right) \right) =\chi _{0}(\Delta_c)+\chi _{1}(\Delta_c)\Omega _s\cos S(t),
\end{equation}
which has a time-invariant part,
\begin{equation}
	\chi _0\left( \Delta _c \right) =\frac{2N_0\mu _{12}^{2}}{{\hbar }\varepsilon _0}\frac{C_{C-DC}}{C_C(\Omega_L)^2},
\end{equation}
and a time variant part coefficient,
\begin{equation}
	\chi _1\left( \Delta _c \right) =\frac{2N_0\mu _{12}^{2}}{{\hbar }\varepsilon _0}\frac{C_{C-AC}}{C_C(\Omega_{L}) ^2}.
\end{equation}

\textbf{(P3): Sensitivity Maximization in General Case via Coupling Laser Detuning}
\begin{equation}
	\Delta _{c}^{*}=\underset{{\Delta _c}}{\text{argmax }}|\chi_1(\Delta_c)|.
\end{equation}

We numerically characterize the frequency detuning effect on the weak signal response in Section IV. 

\section{Numerical Results}\label{sec.numerical}
We adopt some parameters from the experimental setups in \cite{jing2020atomic}. Cs atoms were filled in a vapour cell at room temperature. The cell contained ground-state atoms at a total density of $N_0=4.89 \times 10^{10}$ cm$^{-3}$. For Rydberg atom EIT, the residence time in the Rydberg state is small. Under typical conditions, the population in Rydberg states is $~0.0001$. Only a small distribution of atomic velocity classes in the vapor cell are selected by the EIT lasers, roughly $~1/100$ for the scheme shown in Fig. \ref{fig.level} \cite{fan2015atom}. The effective atoms density for atomic vapor is approximated as $N_{0,\text{eff}}\approx 4.89 \times 10^{8}$ cm$^{-3}$. The four-level configuration in Fig. \ref{fig.level} using four states in a Cs atom: $|1\rangle \rightarrow 6S_{1/2}, F=4$; $|2\rangle \rightarrow 6P_{3/2}, F=5$; $|3\rangle \rightarrow 47D_{5/2}$; and $|4\rangle \rightarrow 48P_{3/2}$. States $|2\rangle$, $|3\rangle$, and $|4\rangle$ have the inverse lifetimes $\gamma_2=2\pi\times 5.2$ MHz, $\gamma_3=2\pi \times 3.9$ kHz, and $\gamma_{4}=2\pi\times 1.7$ kHz. In this section, we ignore the atoms collision effect, $\gamma_c=0$, the effective Rabi frequencies for the transitions $|1\rangle \rightarrow |2\rangle$ and $|2\rangle \rightarrow |3\rangle$ are $\Omega_p=2\pi\times 5.7$ MHz, $\Omega_c=2\pi\times 0.97$ MHz, respectively. The frequency intervals between the hyperfine states $6P_{3/2}$ are $151.2$ MHz ($F=2\rightarrow F=3$), $201.2$ MHz ($F=3\rightarrow F=4$), and $251.1$ MHz ($F=4\rightarrow F=5$) \cite{steck2003cesium}. In the case of resonance, the system is locked to a specific hyperfine state. We consider a detuning range from $-50$ MHz to $50$ MHz for laser frequency and another range from $-100$ MHz to $100$ MHz for electric field frequency in numerical results. 

\begin{figure}[htbp]
	\centering
	\includegraphics[width = 3.5 in]{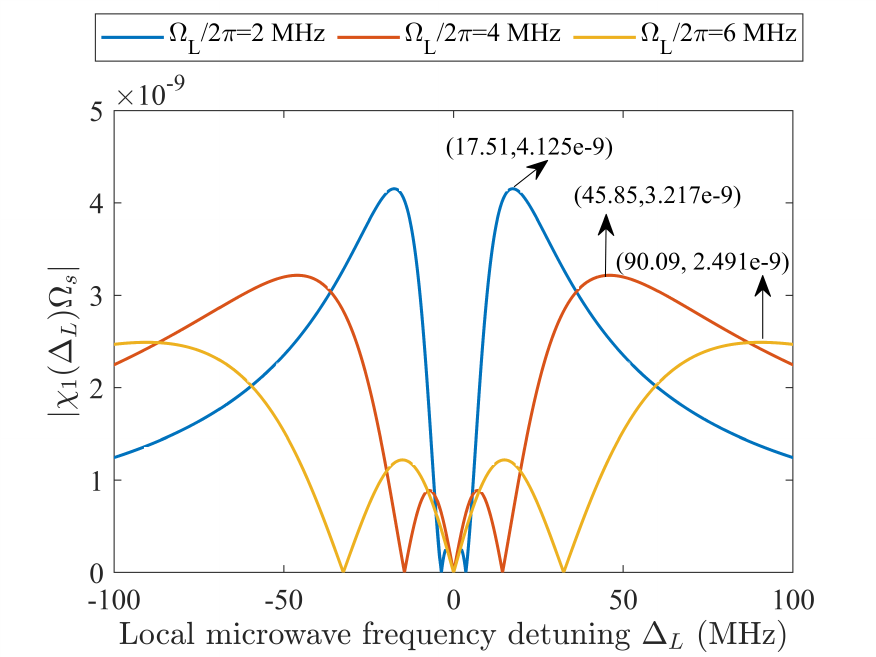}
	\caption{Conversion coefficient $|\chi_1(\Delta_L)\Omega_s|$ with respect to local microwave detuning $\Delta_{L}$ in the general case, $\Omega_s=2\pi\times 10^5$ Hz \textbf{(P1)}.}
	\label{fig.optimal l}
\end{figure}

\textbf{Local microwave frequency detuning case:} The relationship between $\Delta_L$ and conversion coefficient $|\chi_1{\Delta_L}\Omega_s|$ is shown in Fig. \ref{fig.optimal l}. The maximum values of $|\chi_1{\Delta_L}\Omega_s|$ at $\Omega_{L}/2\pi=2$ MHz, 4 MHz, 6 MHz are $4.125\times 10^{-9}$, $3.217\times10^{-9}$, $2.491\times10^{-9}$, respectively. It can be found that weaker local microwave electric field strength corresponds to larger conversion coefficient maximum. At the same time, the detuning value at the maximum point will also become smaller.

\begin{figure}[htbp]
	\centering
	\includegraphics[width = 3.5 in]{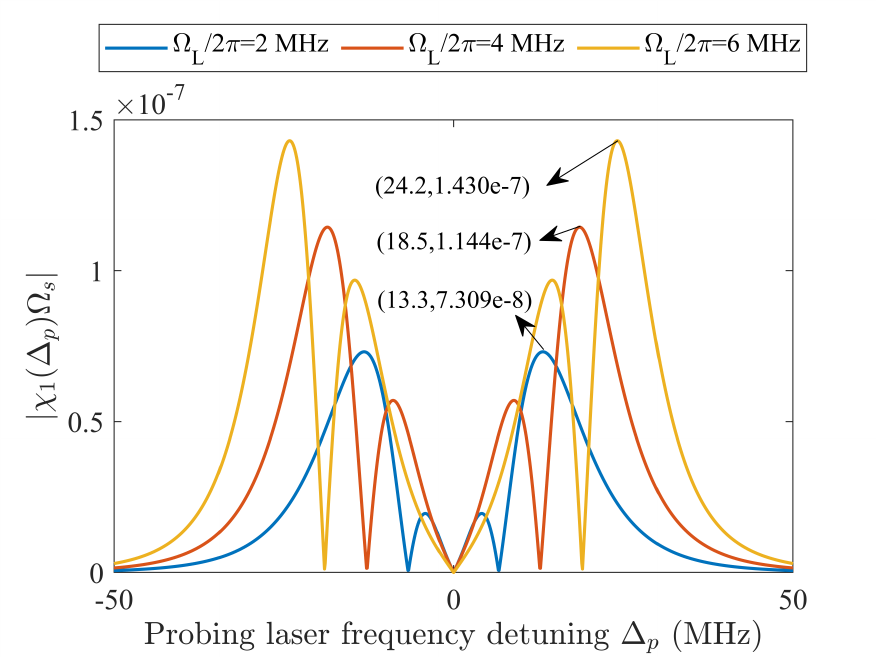}
	\caption{Conversion coefficient $|\chi_1(\Delta_p)\Omega_s|$ with respect to probing laser detuning $\Delta_{p}$ in the general case, $\Omega_s=2\pi\times 10^5$ Hz \textbf{(P2)}.}
	\label{fig.optimal p}
\end{figure}

\textbf{Probing laser frequency detuning case:} The relationship between $\Delta_p$ and conversion coefficient $|\chi_1{\Delta_p}\Omega_s|$ is shown in Fig. \ref{fig.optimal p}. The maximum values of $|\chi_1{\Delta_p}\Omega_s|$ at $\Omega_{L}/2\pi=2$ MHz, 4 MHz, 6 MHz are $7.309\times 10^{-8}$, $1.144\times 10^{-7}$, $1.430\times 10^{-7}$, respectively. In contrast to local microwave frequency detuning case, stronger local microwave electric field strength corresponds to larger conversion coefficient maximum in this case. Besides, we can obtain bigger conversion coefficient by setting optimal probing laser frequency. 

\begin{figure}[htbp]
	\centering
	\includegraphics[width = 3.5 in]{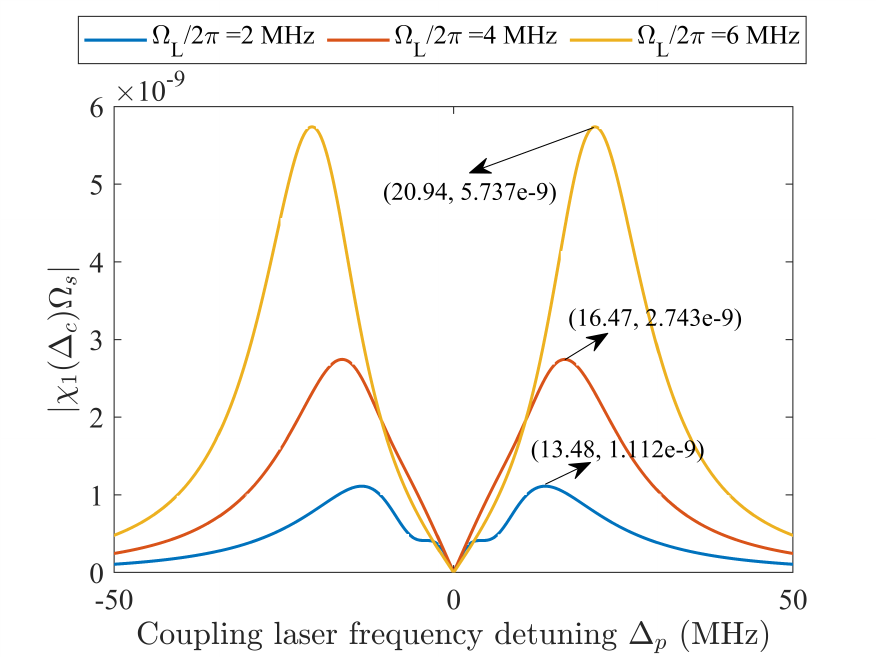}
	\caption{Conversion coefficient $|\chi_1(\Delta_c)\Omega_s|$ with respect to coupling laser detuning $\Delta_{c}$ in the general case, $\Omega_s=2\pi\times 10^5$ Hz \textbf{(P3)}.}
	\label{fig.optimal c}
\end{figure}

\textbf{Coupling laser frequency detuning case:} The relationship between $\Delta_c$ and conversion coefficient $|\chi_1{\Delta_c}\Omega_s|$ is shown in Fig. \ref{fig.optimal c}. The maximum values of $|\chi_1{\Delta_p}\Omega_s|$ at $\Omega_{L}/2\pi=2$ MHz, 4 MHz, 6 MHz are $1.112\times 10^{-9}$, $2.743\times 10^{-9}$, $5.737\times 10^{-7}$, respectively. This case has similar gain than local microwave frequency detuning case.

\section{Sensitivity Estimation}
\subsection{Calculation}

The power sensitivity of a conventional receiver is usually expressed by the minimum power received by the antenna connector to guarantee certain throughput or BER (bit-error-rate) requirements. Assuming that the length of the atomic vapor cell is $L$, the probing laser beam diameter is $d$, the effective signal acquisition area $A_\text{eff}$ equals to L times d. The equivalent signal power sensitivity is given by
\begin{equation}\label{eq.calculation}
	P_{\min\text{\,\,}}=\left< |\vec{S}| \right> \cdot A_{\text{eff}}=\frac{1}{2}\varepsilon _0cE_{\min\text{\,\,}}^{2}\cdot A_{\text{eff}},
\end{equation}
where $\varepsilon_0$ and $c$ are the dielectric constant and speed of light in vacuum, $E_{min}$ is the minimum detectable electric field strength, and $\vec{S}$ is the Poynting vector of electromagnetic field. 

\subsection{Atomic shot noise limit sensitivity}
Theoretically, the minimum detectable electric field strength corresponding to the shot noise limit is given by \cite{fan2015atom}
\begin{equation}
	E_{\min}=\frac{h}{\mu _{\text{E}}T\sqrt{N}},
\end{equation}
where $T$ is the integration time, $N$ is the atom involved in the measurement in $T$ time, and $h$ and $\mu_E$ represent Planck constant and transition dipole momentum coupled electric field, respectively. Considering the limit of the actual EIT (electromagnetic induced transparency) process on time $T$ ($T$ cannot be longer than the $T_2$ time of the EIT process), a more useful form can be obtained. The measurement that occurs within $T^{'}$ can be understood as $N = N_{at}T^{'}/T_2$, where $N_{at}$ is the number of atoms involved in the measurement. The minimum detectable electric field strength of the EIT process can be expressed as
\begin{equation}
	E_{\min}=\frac{h}{\mu_{\text{E}}\sqrt{N_{\text{at}}T^{'}T_2}}.
\end{equation}

Taking $T^{'} = 1$ s, we can get the time-normalized electric field detection sensitivity as
\begin{equation}
	\frac{E_{\min}}{\sqrt{\text{Hz}}}=\frac{h}{\mu _{\text{E}}\sqrt{N_{\text{at}}T_2}}.
\end{equation}

The Rydberg atomic heterodyne receiver has achieved great improvement in sensitivity, the system electric field measurement sensitivity obtained in the experiment reached $55$ nV$\cdot$cm$^{-1}$Hz$^{-1/2}$ \cite{jing2020atomic}. Using the power sensitivity calculation method in Eq. (\ref{eq.calculation}), the power sensitivity of the system is approximately -144.7 dBm$\cdot$Hz$^{-1}$ based on the data provided in the literature. Based on the Rydberg atomic heterodyne receiver, the optimal electric field measurement sensitivity obtained by combining coupled laser frequency detuning is about $12.5$ nV$\cdot$cm$^{-1}$Hz$^{-1/2}$ \cite{cai2022sensitivity}, which corresponds to a power sensitivity about $-157.5$ dBm$\cdot$Hz$^{-1}$ (assuming that the vapor cell size and laser beam diameter parameters are consistent with the parameters in the work of Jing et al. \cite{jing2020atomic}).
The detection sensitivity of the Rydberg atomic heterodyne receiver is mainly limited by the thermal noise of electronic devices, and the technical noise caused by atomic transition broadening and laser linewidth. The atomic transition broadening can be reduced by amplifying the probing laser and coupling laser beam waist. Replacing balanced detection with homodyne or heterodyne detection can reduce photodetector noise. Using a more stable laser source can eliminate laser frequency noise. These factors can be combined to allow the system to approximate the quantum projection noise limit of atomic sensors, about $700$ pV$\cdot$cm$^{-1}$Hz$^{-1/2}$ \cite{jing2020atomic}. The power detection sensitivity corresponding to the quantum projection noise limit is about $-182.6$ dBm$\cdot$Hz$^{-1}$.

\subsection{Photon shot noise limit}
As described in Section V.A, intensity-based readout methods have achieved great improvement in sensitivity, but are still far from the theoretical limit. In this paper, the performance of the zero-difference readout method is evaluated, and the relationship between the phase conversion coefficient and frequency detuning under different local oscillator electric field strengths is obtained. Referring to the results of the phase noise, the corresponding minimum detectable electric field strength can be given to calculate the power sensitivity.

We model an ideal measurement by assuming a photon-shot-noise limited homodyne measurement of the acquired phase $\phi$ on the transmitted probe light, where $\phi$ is calculated by Eq. (\ref{eq.delta_phi}). The noise density $ phi_N$ in measuring this quantity for a coherent state is defined as
\begin{equation}
	\phi _N=\frac{2\mu _{21}}{d\Omega _p}\sqrt{\frac{k_p}{h\varepsilon _0}},
\end{equation}
where detection efficiency assumed to be unity in this work. 

The SNR (signal-to-noise ratio) density is given by
\begin{equation}
	\frac{\Delta \phi}{\phi _N}=\frac{k_pLd\Omega _p\Omega _s}{8\mu _{21}}\sqrt{\frac{h\varepsilon _0}{k_p}}\chi _1\left( \Delta \right).
\end{equation}
Assume that readout requirement is $0$ dB, the minimum detectable electric field is given by
\begin{equation}\label{eq.emin}
	\frac{E_{\min}}{\sqrt{\text{Hz}}}=\frac{4\mu _{21}}{\pi Ld\mu _E\Omega _p\chi _1\left( \Delta \right)}\sqrt{\frac{h}{k_p\varepsilon _0}}.
\end{equation}

Numerical results in Fig. \ref{fig.optimal l}, Fig. \ref{fig.optimal p}, and Fig. \ref{fig.optimal c} show that the largest conversion coefficient is obtained in the probing laser frequency detuning case with $\Omega_L=2\pi\times 6$ MHz, $\chi_1(\Delta_p) = 1.430\times 10^{-7}$. According to Eq. (\ref{eq.emin}), the electric field detection sensitivity is $0.185$ nV$\cdot$cm$^{-1}$Hz$^{-1/2}$, and the corresponding power detection sensitivity is $-194.1$ dBm$\cdot$Hz$^{-1}$. The phase-based readout method improves the sensitivity with $36.6$ dB gain. It is important to note that we consider the sensitivity improvement of three frequency components detuning independently for three arbitrarily selected local oscillator electric field strengths. The results obtained are not theoretically optimal sensitivity of the proposed method in this work.



\section{Conclusion}
Based on the atomic superheterodyne receiver and Mach-Zehnder interferometer, we evaluate the detection sensitivity of phase-based readout method. We have derived the steady-state solutions of the density matrix of the Rydberg atomic system operating with frequency detuning. Given the input microwave signal, we have established an optimization problem of the linear susceptibility of atomic medium to maximize the amplitude of the output phase signal. We have numerically analyzed the effects of electric field, probing laser, and coupling laser frequency detuning on the optimization target. Numerical results show that the detection sensitivity can be improved by setting appropriate frequency detuning, especially for probing laser. The photon shot noise limit sensitivity of the phase-based readout method has 36.6 dB gain compared with that of experimental result of the intensity-based readout method.

\bibliographystyle{./IEEEtran}
\bibliography{./mybib}

\end{document}